\documentclass[twocolumn,amsmath,amssymb,longbibliography,superscriptaddress,aps,pra,10pt]{revtex4-2}

\usepackage[separate-uncertainty = true,multi-part-units=single]{siunitx}
\usepackage{graphicx}
\usepackage[english]{babel}
\usepackage[utf8]{inputenc} 
\usepackage{ctable} 
\usepackage{booktabs}
\usepackage{wasysym}
\usepackage{color, colortbl}
\usepackage{german}
\usepackage[colorlinks=true,allcolors=blue]{hyperref}

\newcommand{\Iso}[2]{$^{#2}${#1}}
\newcommand{\Ar}[1]{$^{#1}$Ar}
\newcommand{\Kr}[1]{$^{#1}$Kr}
\newcommand{\C}[1]{$^{#1}$C}
\newcommand{\Cl}[1]{$^{#1}$Cl}
\newcommand{\dO}{$\delta^{18}$O }

\definecolor{Lightblue}{rgb}{0.85,0.85,0.95}
\definecolor{Lightred}{rgb}{0.95,0.9,0.9}

\begin{document}

\title{\Kr{81} dating at the Guliya ice cap, Tibetan Plateau}
\author{Lide Tian}
\email{ldtian@ynu.edu.cn}
\affiliation{Institute of International Rivers and Eco-security, Yunnan University, Kunming 650500, China.}
\affiliation{CAS Center for Excellence in Tibetan Plateau Earth Sciences, Chinese Academy of Sciences, Beijing 100101, China.}
\affiliation{College of Resource and Environment, University of Chinese Academy of Sciences, Beijing, 100190, China.}
\affiliation{Yunnan Key Laboratory of International Rivers and Transboundary Eco–security, Yunnan University, Kunming 650091, China.}

\author{Florian Ritterbusch}
\author{Ji-Qiang Gu}
\author{Shui-Ming~Hu}
\author{Wei Jiang}
\affiliation{Hefei National Laboratory for Physical Sciences at the Microscale, CAS Center for Excellence in Quantum Information and Quantum Physics, University of Science and Technology of China, Hefei 230026, China}

\author{Zheng-Tian Lu}
\email{ztlu@ustc.edu.cn}
\affiliation{Hefei National Laboratory for Physical Sciences at the Microscale, CAS Center for Excellence in Quantum Information and Quantum Physics, University of Science and Technology of China, Hefei 230026, China}

\author{Di Wang}
\affiliation{Institute of International Rivers and Eco-security, Yunnan University, Kunming 650500, China.}

\author{Guo-Min Yang}
\affiliation{Hefei National Laboratory for Physical Sciences at the Microscale, CAS Center for Excellence in Quantum Information and Quantum Physics, University of Science and Technology of China, Hefei 230026, China}

\begin{abstract}
We present radiometric \Kr{81} dating results for ice samples collected at the outlets of the Guliya ice cap in the western Kunlun Mountains of the Tibetan Plateau. This first application of \Kr{81} dating on mid-latitude glacier ice was made possible by recent advances in Atom Trap Trace Analysis, particularly a reduction in the required sample size down to \SI{1}{\mu L\ STP} of krypton. Eight ice blocks were sampled from the bottom of the glacier at three different sites along the southern edges. The \Kr{81} data yield upper age limits in the range of 15\hspace{0.05cm}-\hspace{0.05cm}\SI{74}{ka} (90\% confidence level). This is an order of magnitude lower than the ages exceeding \SI{500}{ka} which the previous \Cl{36} data suggest for the bottom of the Guliya ice core. It is also significantly lower than the widely used chronology up to \SI{110}{ka} established for the upper part of the core based on \dO in the ice. 
\end{abstract}


\maketitle

\section{Introduction}

\begin{figure*}
\centering
\noindent \includegraphics[width=15cm]{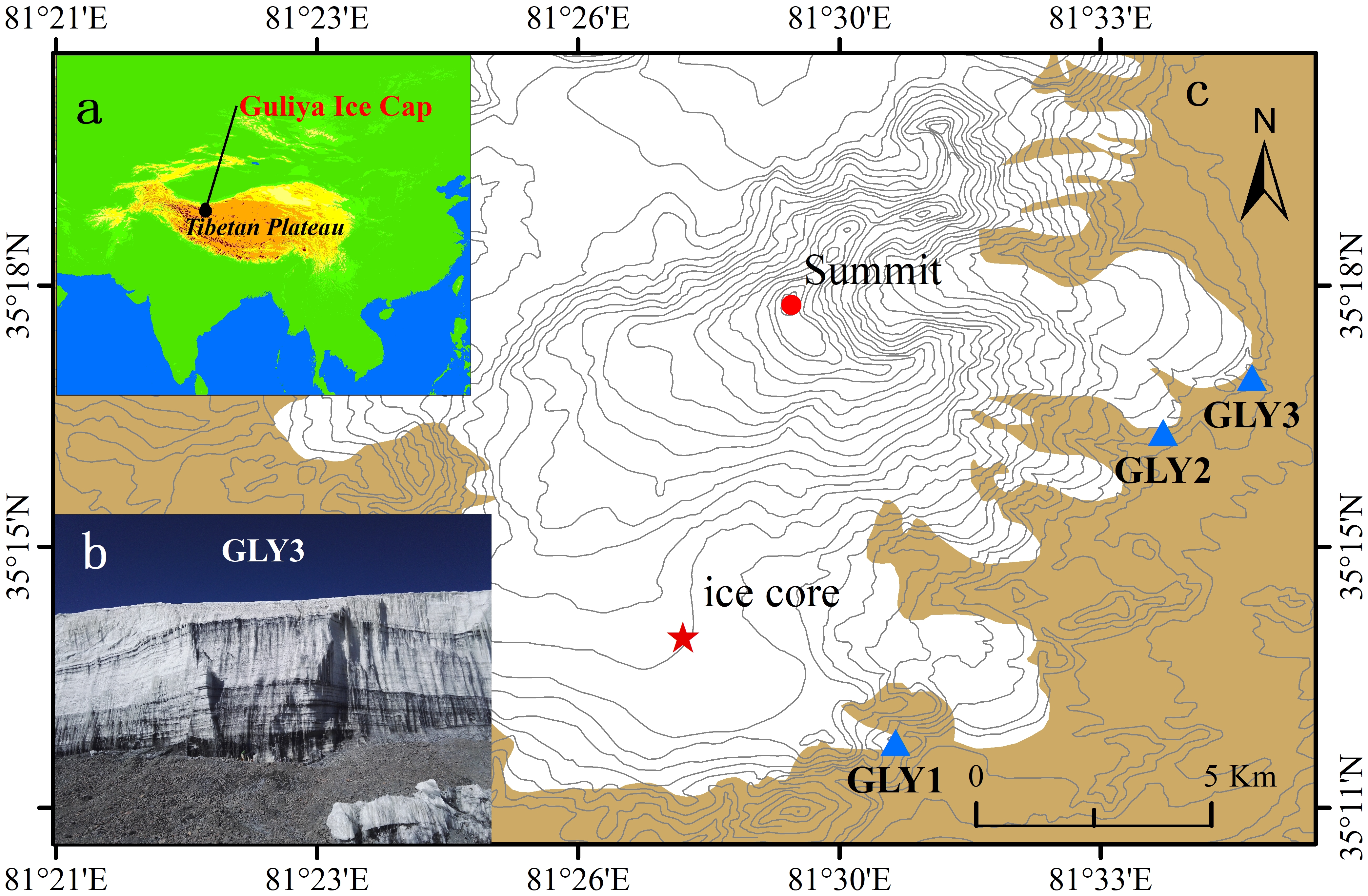}
\caption{(a) Location of the Guliya ice cap on the Tibetan Plateau; (b) photograph showing the glacier cliff ($\sim$\hspace{0.05cm}\SI{20}{m} tall) at sampling site GLY3; (c) Sampling sites GLY1, GLY2 and GLY3 for bottom ice of the Guliya ice cap during 2015\hspace{0.05cm}-\hspace{0.05cm}2017. The red dot marks the summit (\SI{6710}{m} a.s.l.) and the red star the location of the Guliya ice core (GIC1992) drilling site (\SI{6200}{m} a.s.l.) from 1992 \cite{Thompson1997}.  }
\label{fig:guliya_sampling}
\end{figure*}

Alpine ice cores in the mid- and low-latitude regions provide high-resolution records of past climate and environment. High rates of ice accumulation and melting are responsible for the relatively short history of ice core records on the Tibetan Plateau as compared to the polar regions. Longer ice cores and older ice are being sought on the Tibetan Plateau for the purpose of extending the climate history in this region. The Malan and Puruogangri ice cores in the central Tibetan Plateau \cite{Thompson2006, Wang2003} and the Dasuopu ice core in the middle of the Himalayas \cite{Thompson2000, Yao2002} provide records of the past several thousand years. Samples from the bottom of the Dunde ice core in the northeastern Tibetan Plateau were first interpreted to be glacial-stage ice \cite{Thompson1989}, but later proved to be a Holocene deposit \cite{Thompson2005}. The longest (\SI{308.6}{m}) ice core and the oldest bedrock ice so far discovered on the Tibetan Plateau is from the Guliya ice cap in the western Kunlun Mountains \cite{Yao1997, Thompson1997}. Developing a chronology for this Guliya ice core (GIC1992 hereafter), as for Tibetan ice cores in general, is challenging. Dating by layer counting is difficult for ice cores from the Tibetan Plateau because the monsoonal type precipitation pattern in this region generates weaker seasonal variation \cite{Hou2004}. For GIC1992 an age scale up to \SI{110}{ka} was established down to \SI{266}{m} depth by comparing the \dO signal with the CH$_4$ record from GISP2 in Greenland. Moreover, the \Cl{36} data suggest that the bottom ice may be older than \SI{500}{ka}. Since then, the GIC1992 record has been widely used as a reference for correlating regional climate signals [e.g. Cheng et al. 2012, Chevalier et al. 2011, Cosford et al. 2008, Hayashi et al. 2009, Mahowald et al. 2011]. 
However, the established Guliya chronology is difficult to reconcile with several recent findings. \cite{Cheng2012} encountered inconsistencies between the \dO record of GIC1992 and the Kesang stalagmite record. Their work suggests that the relationship between \dO and CH$_4$ may be inversed, leading to a shortening of the GIC1992 age scale by a factor of two. Meanwhile, at the Chongce ice cap ($\sim$\hspace{0.05cm}\SI{30}{km} away from the GIC1992 drilling site), luminescence dating provides an upper age limit of \SI{42 \pm 4}{ka} for the basal sediment \cite{Zhang2018}, which is an order of magnitude lower than what the \Cl{36} data suggests for the bottom ice of GIC1992. Moreover, \C{14} dating in combination with ice flow modeling for ice cores from the Chongce ice cap indicates Holocene deposition \cite{Hou2018}, which is consistent with all other Tibetan ice cores except GIC1992. Given the proximity between the Guliya and the Chongce ice cap, these results make it difficult to argue that the large difference in age scale between GIC1992 and the other Tibetan ice cores is due to different local climate conditions in the western Kunlun Mountains \cite{Thompson2005}. All the foregoing findings raise the need for examining the GIC1992 chronology with an independent dating method.

\Kr{81} is a cosmogenic radionuclide with a half-life of \SI{229 \pm 11}{ka}. The \Kr{81} concentration in the atmosphere (isotopic abundance \Kr{81} /Kr $\sim$\hspace{0.05cm}$10^{-12}$) is spatially homogeneous with only small changes over the past 1.5 million years \cite{Buizert2014}. These properties as well as its chemical inertness make it a desirable tracer for groundwater and ice over the age range of \SI{40}{ka} to \SI{1.3}{Ma} \cite{Loosli1969, Lu2014}. Meanwhile, the anthropogenic \Kr{85} (half-life \SI{10.76\pm0.02}{a}), which is mainly produced by nuclear fuel reprocessing, can be used to identify any young ($<$\hspace{0.05cm}\SI{60}{a}) components or contamination of an old sample with modern air \cite{Winger2005}. Development of the analytical method of Atom Trap Trace Analysis (ATTA) has made radiokrypton dating available to the earth science community at large \cite{Jiang2012}. Due to the large required sample size (5\hspace{0.05cm}-\hspace{0.05cm}\SI{10}{\mu L\ STP} of krypton), so far \Kr{81} has been used mainly for dating groundwater while for glacier ice only a demonstration study was conducted on large blue ice samples ($\sim$\hspace{0.05cm}\SI{350}{kg}) from Taylor Glacier, Antarctica \cite{Buizert2014}. Recently, the required sample size for \Kr{81}- and \Kr{85}-analysis has been reduced down to \SI{1}{\mu L\ STP} of krypton, which can be extracted from about \SI{10}{kg} of Antarctic ice (containing $\sim$\hspace{0.05cm}\SI{100}{mL\ STP} air per kg ice) or 20\hspace{0.05cm}-\hspace{0.05cm}\SI{40}{kg} of Tibetan glacier ice (25\hspace{0.05cm}-\hspace{0.05cm}\SI{50}{mL\ STP} air/kg) \cite{Li2011}. This sample size is still too large to re-assess the historic GIC1992 directly, but is sufficient for \Kr{81} dating of samples from the margin sites of the Guliya ice cap, as presented in this work.

\section{Methods}

\subsection{Site description and ice sampling}
Guliya is a large ice cap in the western Kunlun Mountains on the Tibetan Plateau with a total area of about \SI{376}{km^2} \cite{Thompson1997, Yao1997}. Its southern part is of nonsurge type with stationary terminus positions \cite{Yasuda2015}. Remote sensing data show that the glaciers in this region have experienced less change in recent decades compared to other glaciated mountainous regions in western China \cite{Shangguan2017}. The Guliya ice cap even gained mass from 2000\hspace{0.05cm}-\hspace{0.05cm}2015 \cite{Kutuzov2018} primarily due to increasing precipitation in the westerly regime \cite{Yao2012}. Ice core drilling and ground penetrating radar show that the glacier thickness varies from about \SI{50}{m} at the summit to a maximum thickness of \SI{371}{m} at a location \SI{1.5}{km} upstream of the GIC1992 drilling site (Figure \ref{fig:guliya_sampling}) \cite{Kutuzov2018, Thompson1997}. The glacier flows from the summit at \SI{6710}{m} altitude down to the margins at approximately \SI{5500}{m} \cite{Thompson1997} with an average slope of $<$ 3 - \SI{5}{\degree} \cite{Kutuzov2018}. Limited field observation indicates increasing negative surface mass balance going from the equilibrium line altitude of around \SI{6000}{m} to lower elevation sites \cite{Li2019}. The ablation of the ice cap is also characterized by cliff melting at the end of the glacier outlets so that the bottom ice layers become accessible over large sections of the glacier edge.

\subsection{Air extraction from the ice samples}
For \Kr{81} and \Kr{85} analysis, the air trapped in the ice has to be extracted. Prior to extraction, the surface of the ice samples is cleaned to remove any layers or flaky debris that may contain modern air. The ice is then brought out of the cold room and placed in a stainless steel chamber which is thereafter sealed and evacuated for about \SI{30}{min} by scroll pumps through a water trap (stainless steel bellow immersed in ethanol at \SI{-80}{\celsius}). Since during pumping the chamber is constantly being flushed by the water vapor from the sublimating ice, the remaining atmospheric gas in the container is rendered negligible. After evacuation, the chamber is heated by a stove for 60\hspace{0.05cm}-\hspace{0.05cm}\SI{90}{min} (depending on the ice mass) until the ice has completely melted. The gas released from the ice passes through the water trap and is compressed into a sample cylinder. The air content of the ice sample is determined based on the final pressure in the sample cylinder (Table \ref{tab:krypton_results}). Extraction efficiencies higher than 95\% and contamination with modern air below 1\% are typically achieved with this degassing method at a processing time of about 2\hspace{0.05cm}-\hspace{0.05cm}\SI{3}{hours} per sample. More details on the extraction system and procedure are provided in the supporting material.

\subsection{Krypton purification and \Kr{81} measurement}
The extracted gas from the ice samples was sent to the University of Science and Technology of China (USTC) for krypton purification and for ATTA analysis of both \Kr{81} and \Kr{85}. Krypton is separated from the extracted gas using a purification system based on titanium gettering and gas chromatography \cite{Tu2014}, typically yielding krypton purities and recoveries both higher than 90\%. 
The \Kr{81} and \Kr{85} measurements are performed with the latest ATTA instrument at USTC, where individual \Kr{81} and \Kr{85} atoms are selectively laser-cooled and then detected in a magneto-optical trap. The stable and abundant \Kr{83} is also measured for normalization. The resulting \Kr{81}/\Kr{83} and \Kr{85}/\Kr{83} ratios for the sample are compared to the corresponding ratios of a reference krypton gas to derive the \Kr{81} abundance as a percentage of the atmospheric value (pMKr) and the \Kr{85} abundance given in the units of dpm/cc (decay per minutes per cc STP krypton), a convention originating from decay counting. More details on \Kr{81} and \Kr{85} analysis with ATTA can be found in \cite{Jiang2012}.

\section{Results and discussion}
\subsection{\dO results}

\begin{figure}
\centering
\noindent \includegraphics[width=9cm]{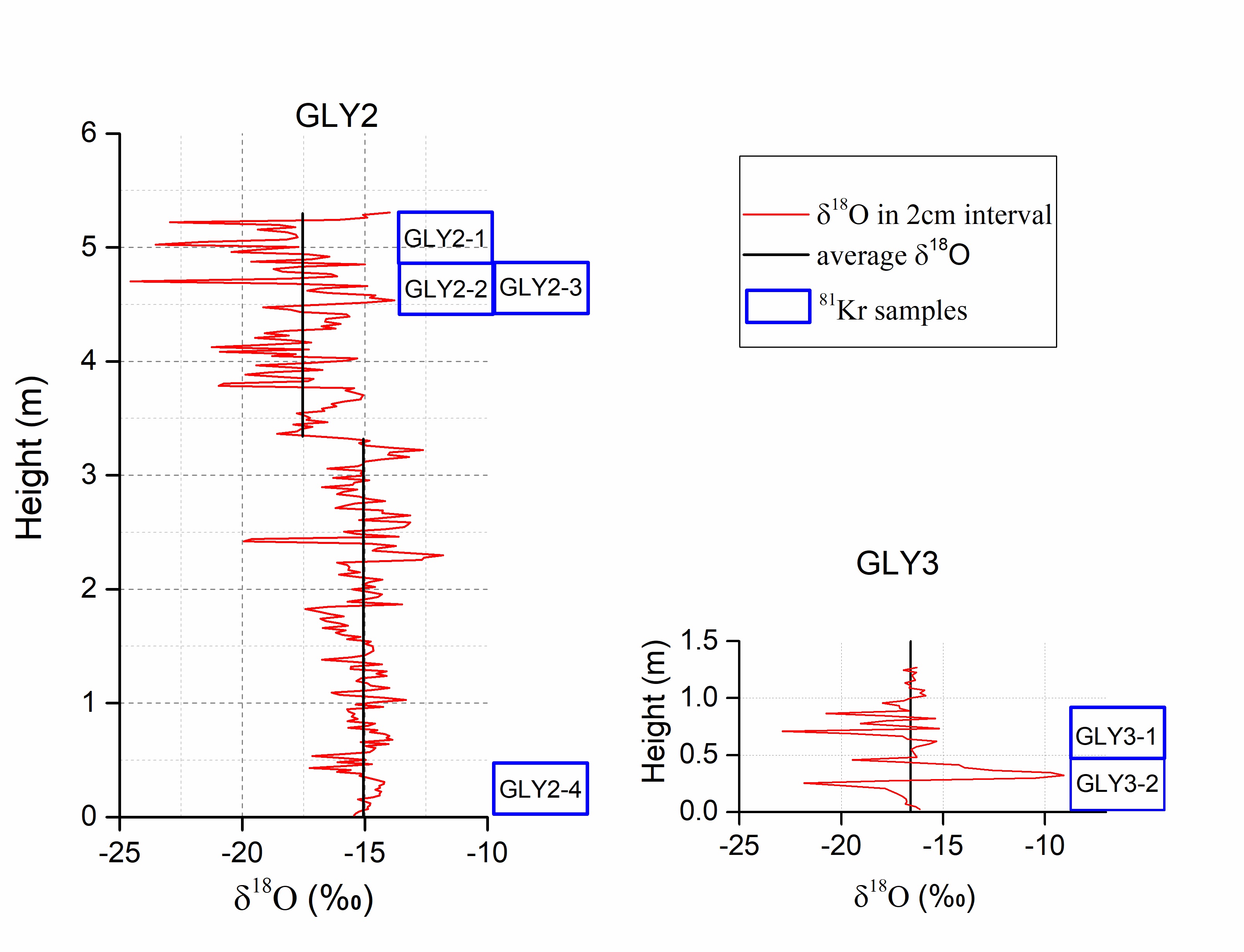}
\caption{Vertical \dO profiles along a \SI{5.3}{m} column at GLY2 and a \SI{1.27}{m} column at GLY3. The boxes show the positions of the \Kr{81}-dated glacier ice samples along the vertical profiles. The zero in height corresponds to the visible bottom of the glacier cliff, but is not necessarily the bedrock as debris may cover the lowermost part of the glacier. The size and the vertical position of the samples are roughly to scale. For GLY2, the \dO data of the lower \SI{3.3}{m} have an average of \SI{-15.0}{\permil} and a standard deviation (std) of \SI{1.1}{\permil} whereas in the upper \SI{2}{m} the average is \SI{-17.5}{\permil} (std=\SI{2.0}{\permil}). For GLY3, the average is \SI{-16.6}{\permil} (std=\SI{2.4}{\permil}).}
\label{fig:dO18_profiles}
\end{figure}

Figure \ref{fig:dO18_profiles} shows the oxygen isotope variation along the \SI{5.3}{m} bottom ice at GLY2 and the \SI{1.27}{m} profile at GLY3. It is difficult to match these short \dO profiles from GLY2 and GLY3 with the \dO record from GIC1992 \cite{Thompson1997}. However, the \dO fluctuations along the profile provide hints whether the ice originates from the bottom or not. The accumulation layers of the glacier rapidly become thinner towards the bottom. The fast fluctuations of the \dO signal is averaged out when the thickness of the layers become less than the \SI{2}{cm} cutting interval. The large fluctuations in the \dO profile at GLY3 as well as in the upper part of GLY2 are comparable to those at the top of GIC1992 \cite{Thompson2018}, suggesting that these samples are not derived from the bottom of the glacier. The samples were collected from the visible lowest part of the glacier cliff, which is not necessarily the lowest part of the ice as the bottom may be covered by debris from the glacier. This explanation is supported by the observation of a large amount of pebbles being deposited in front of the glacier cliff at GLY3. In contrast, the fluctuations in the \dO profile at GLY2 exhibit reduced fluctuations towards the bottom of the glacier. This indicates that GLY2-4, collected at the bottom of the GLY2 profile, is likely close to the very bottom of the glacier ice. The reduced fluctuations in the lower \SI{3.3}{m} of the \dO profile at GLY2 may also result from mixing of ice of different ages due to complex flow leading to averaging of the \dO values. The same mechanism may be responsible for the higher fluctuations in the \dO profile at GLY3 and at the top of GLY2 (e.g. if layers with higher \dO values are transported next to layers with lower \dO values) although no stratigraphic disturbance has been observed at the three sampling sites. \\
It is difficult to match the \dO records from this study to the one from GIC1992 because of the high ambiguity in matching the excursions and because the ice at GLY2 and GLY3 originates from a different accumulation zone than the ice at the GIC1992 drilling site. At the height of \SI{3.3}{m} the \dO record at GLY2 exhibits a shift in the mean from \SI{-17.5}{\permil} to \SI{-15}{\permil} and below that the standard deviation is reduced from \SI{2}{\permil} to \SI{1.1}{\permil} (Figure \ref{fig:dO18_profiles}). This behavior is similar for the \dO signal of GIC1992 with the difference that the mean \dO value at the bottom \SI{40}{m} is higher than the bottom \SI{3.3}{m} at GLY2 by about \SI{2}{\permil}. This is likely due to the altitude difference of the accumulation zones of the ice at GLY2 and GIC1992.

\subsection{Air content}
The measured air contents in the ice samples are listed in Table \ref{tab:krypton_results}. They vary from \SI{32}{mL\ STP/kg} to \SI{59}{mL\ STP/kg}, which is typical for Himalayan ice cores \cite{Hou2007, Li2011} and significantly lower than the air content of Antarctic ice, typically ranging between 100\hspace{0.05cm}-\hspace{0.05cm}\SI{120}{mL\ STP/kg} \cite{Buizert2014, Raynaud1979}, or that of Greenland ice at 80\hspace{0.05cm}-\hspace{0.05cm}\SI{100}{mL\ STP/kg} \cite{Raynaud1997}. This is due to the lower air pressure at high elevation (5500-\SI{6700}{m}) of the deposition site and the higher temperature compared to Antarctica \cite{Eicher2016, Martinerie1992}. We deliberately collected the samples from ice layers with visibly high bubble content and avoided those with transparent ice which are likely layers of re-frozen meltwater.

\begin{table*}
\caption{Compilation of the \Kr{81} and \Kr{85} results. The \Kr{81} abundance is reported in units of pMKr (percent Modern Krypton). The atmospheric level is \SI{100}{pMKr}. The \Kr{85} abundance is reported in the units of dpm/cc (decays per minute per cc STP of krypton). The errors are 1$\sigma$ standard deviations whereas upper limits are reported for a 90\% confidence level. \\}
\centering
\def\arraystretch{1.3}
\begin{tabular}{c|c|c|c|c|c|c|c}
\toprule 
\rowcolor{Lightblue}
\hspace{0.4cm} sample 	\hspace{0.4cm}	 		& \hspace{0.3cm}  note   \hspace{0.3cm}  		& \hspace{0.2cm}weight  \hspace{0.2cm}      &\hspace{0.1cm} Air content 	\hspace{0.1cm}	&  \hspace{0.1cm}Krypton \hspace{0.1cm}		& \hspace{0.5cm} \Kr{85} \hspace{0.5cm}	&\hspace{0.4cm}\Kr{81} \hspace{0.4cm} 	& \hspace{0.2cm} \Kr{81}-age\hspace{0.2cm} \\
\rowcolor{Lightblue}
	      		 		& 		    		&  kg			  & mL STP/kg    	& $\mu$L STP	& dpm/cc 	&pMKr		&  ka				\\
\midrule

GLY1-1	& Cave 		&   34    		  &  52  				&  1.4 			& $< 1.5$	& \num{97 \pm 7}	& $<52$ \\
GLY1-2	& Cave 		&   28    		  &  46  				&  1.3 			& \num{1.6 \pm 0.2}	& \num{106 \pm 6}	& $<15$ \\ \rowcolor{Lightred}

GLY2-1	& Cave 		&   53    		  &  37  				&  1.1 			& \num{6.1 \pm 1.8}	& \num{93 \pm 7}	& $<74$ \\ \rowcolor{Lightred}
GLY2-2	& Cave 		&   69    		  &  41  				&  2.5 			& \num{2.5 \pm 0.2}	& \num{97 \pm 5}	& $<39$ \\ \rowcolor{Lightred}
GLY2-3	& Surface  	&   52  		  &  45  				&  1.7 			&  \num{0.7 \pm 0.2}	& \num{97 \pm 5}	& $<39$ \\ \rowcolor{Lightred}
GLY2-4	& Surface 	&   30    		  &  32  				&  0.7 			& $< 0.4$	& \num{104 \pm 7}	& $<25$ \\

GLY3-1	& Surface 	&   43    		  &  29  				&  1.8 			&  \num{1.0 \pm 0.2}	& \num{93 \pm 5}	& $<58$ \\ 
GLY3-2	& Surface 	&   36    		  &  50  				&  1.4 			&  \num{4.1 \pm 0.4}	& \num{98 \pm 6}	& $<45$ \\ \rowcolor{Lightred}

Lhasa-Air1	& Surface 	&   43    		  &  29  				&  1.8 			& \num{75 \pm 2}	& - 	&- \\ \rowcolor{Lightred}
Lhasa-Air2	& Surface 	&   36    		  &  50  				&  1.4 			& \num{76 \pm 3}	& - 	& - \\

\bottomrule
\end{tabular}
\label{tab:krypton_results}
\end{table*}

\subsection{\Kr{85} and \Kr{81} results}
The measured \Kr{81} and \Kr{85} abundances for the eight glacier ice samples as well as two air samples of Lhasa are listed in Table \ref{tab:krypton_results}. As described above, the \Kr{85} in the atmosphere has almost exclusively been produced anthropogenically in the past \SI{60}{years}. Therefore, any sample older than that should have a vanishing \Kr{85} abundance. Five of the eight samples have \Kr{85} activity levels below 3\% of the Lhasa air value (Table \ref{tab:krypton_results}), whereas GLY2-1, GLY2-2 and GLY3-2 have \Kr{85} values corresponding to about 8\%, 3\% and 5\%, respectively. Air leaks are thoroughly investigated on instruments used in the degassing, purification and ATTA measurement, leading to the conclusion that contamination of modern air during these processes is below 1\%. It thus seems more likely that modern air had already entered the ice prior to sampling, e.g. by cracking/melting and refreezing, as has been observed in earlier works on glacier ice close to the surface of margin sites \cite{Craig1990, Buizert2014}. Since there is no obvious correlation between \Kr{85} and whether the sample is from the surface or from a cave, potential contamination processes at the very front of the glacier ice cliff do not seem to be responsible for that. 
Since the measured \Kr{81} abundances are close to the modern value of \SI{100}{pMKr}, contamination of modern air at these low concentrations does not affect the reported \Kr{81} abundances within the given precisions. For all samples they are consistent with modern atmospheric \Kr{81} abundance within $1\sigma$, except for GLY3-1 which still lies within a 2$\sigma$ error. We translate the measured relative \Kr{81} abundances into \Kr{81}-ages using the Feldman-Cousins method \cite{Feldman1998}. As the \Kr{81} abundances are close to modern, this method yields upper age limits (90\% confidence level) for the individual samples that range between 15\hspace{0.05cm}-\hspace{0.05cm}\SI{74}{ka}.

\subsection{Implication for the Guliya ice core chronology}
The obtained results for \Kr{81} and \dO of the Guliya margin samples allow for a discussion in the context of the results from GIC1992 \cite{Thompson1997} (see introduction). The \Kr{81} measurements do not show evidence for ice older than \SI{74}{ka} at the bottom of the sampled margin sites of the Guliya ice cap. For the samples from GLY1, where the ice from GIC1992 is expected to outcrop \cite{Kutuzov2018}, the upper limits for the \Kr{81} age do not exceed \SI{52}{ka}. For GLY2-4, whose \dO profile exhibits bottom ice characteristics, the \Kr{81} results provide an upper age limit of only \SI{25}{ka}. The obtained upper age limits do not necessarily rule out the existence of older ice somewhere else in the Guliya ice cap. It is possible that the old ice at the bottom of GIC1992 is frozen to the bedrock and does not flow out to the margin sites. However, radar measurements indicate that the ice at the bottom of GIC1992 does flow and is not trapped at the bedrock \cite{Kutuzov2018}. A further explanation is that the stratigraphy of the glacier ice is folded when travelling from the GIC1992 drilling site to the margin, such that the old ice may not be at the bottom. No evidence for folding was observed at the glacier terminals, which exhibit clear horizontal layer structures, but folding on intermediate distance scales may have occurred. Yet another possibility is that the bottom \SI{100}{m} of GIC1992, which are supposedly older than \SI{50}{ka}, are rapidly thinning towards the outlet of the glacier, and therefore may be contained in a much smaller vertical extent at the very bottom of the glacier cliff. Since the samples at GLY1 were taken in about \SI{2}{m} height above bedrock, they may not reach into this old bottom section. However, measurements of the mass balance and the glacier surface velocity \cite{Li2019, Chadwell2017} indicate that a large fraction of the upper glacier layers is lost when flowing from the equilibrium line altitude to the edge of the glacier at GLY1 where the remaining glacier cliff is about \SI{10}{m} in height. Therefore, it does not seem likely that the bottom \SI{100}{m} at the GIC1992 drilling site are thinning to below our sampling height about \SI{2}{m} above bedrock at GLY1.

\section{Conclusions and Outlook}

Radiometric \Kr{81} dating has been used to determine the age of bottom ice samples at the Guliya ice cap. Eight ice blocks, each weighing 28\hspace{0.05cm}-\hspace{0.05cm}\SI{69}{kg}, were collected at three different outlets of the glacier, and analyzed for \Kr{81} using the Atom Trap Trace Analysis method. The \Kr{81} results yield upper limits in the range of 15\hspace{0.05cm}-\hspace{0.05cm}\SI{74}{ka}, which is an order of magnitude lower than previously suggested by \Cl{36} dating of the Guliya ice core and also significantly lower than the Guliya chronology reaching up to \SI{110}{ka} based on \dO measurements. After results from the Kesang stalagmite cave ($\sim$\hspace{0.05cm}\SI{860}{km} distance to the Guliya ice cap) and the Chongce ice cap ($\sim$\hspace{0.05cm}\SI{30}{km} distance), the \Kr{81} data in this work (obtained directly from bottom samples of the Guliya ice cap) represent yet another result that calls for further dating measurements to check the established Guliya chronology. Measurements of \C{14}, \Cl{36}, \Iso{Be}{10}, $\delta^{18}$O$_{\text{atm}}$ and argon isotope ratios are planned for a new Guliya ice core that has been drilled in 2015 close to the location of the 1992 Guliya core drilling site [Thompson et al. 2018]. Meanwhile, at the USTC laboratory, work is in progress to further reduce the sample size required for \Kr{81} analysis so that bottom samples from a Guliya ice core can be measured directly.


\begin{acknowledgments}
This work is funded by National Natural Science Foundation of China (41530748), the National Key Research and Development Program of China (2016YFA0302200) and the Chinese Academy of Sciences (XDB21010200). We thank Lili Shao and Cheng Wang from the Institute of Tibetan Plateau Research for their assistance in the ice degassing and Lei Zhao from USTC for purifying the krypton samples.  \\ \\

\textit{An edited version of this paper was published by AGU. Copyright 2019 American Geophysical Union.} \\

\textit{Tian, L., Ritterbusch, F., Gu, J.‐Q.,
Hu, S.‐M., Jiang, W., Lu, Z.‐T., et al.
(2019). \Kr{81} dating at the Guliya ice cap,
Tibetan Plateau. Geophysical Research
Letters, 46. https://doi.org/10.1029/
2019GL082464.}
\end{acknowledgments}

\newpage

\appendix*

\renewcommand{\thefigure}{S\arabic{figure}}

\setcounter{figure}{0}

\section*{Supporting material}

\subsection*{Photographs showing the sampling sites}
\begin{figure}[ht!]
\centering
\noindent \includegraphics[width=8cm]{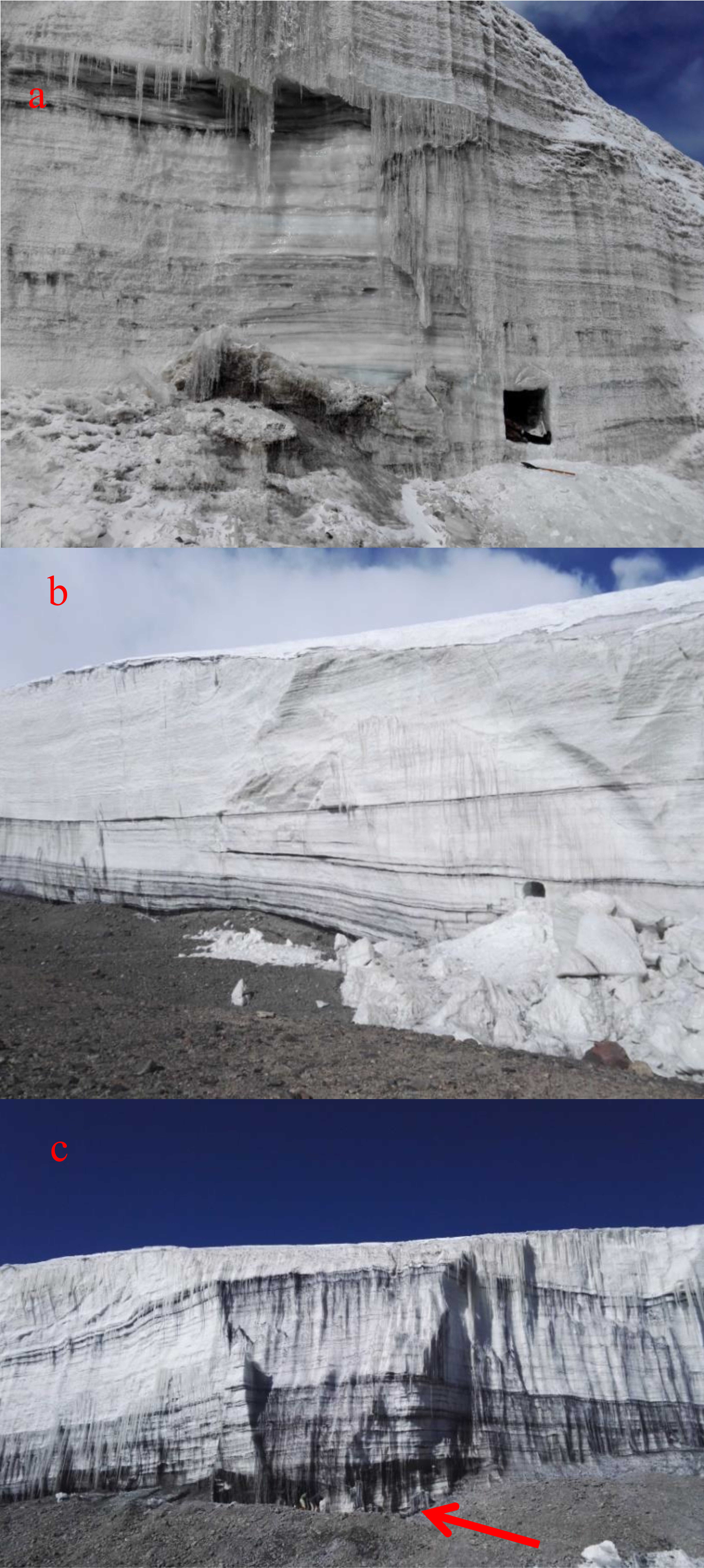}
\caption{Photographs showing the sampling sites GLY1 (a), GLY2 (b) and GLY3 (c) at the Guliya ice cap. In (a) and (b) the caves that have been dug for sampling are shown. They are about \SI{1.5}{m} in height. In (c) the sampling location at the surface is shown. The glacier cliff is about \SI{20}{m} tall.}
\label{fig:photographs}
\end{figure}

Photographs of the glacier cliffs at the three different sampling sites are shown in Figure \ref{fig:photographs}.

\subsection*{Extraction of air from the ice samples}
\begin{figure}
\centering
\noindent \includegraphics[width=8cm]{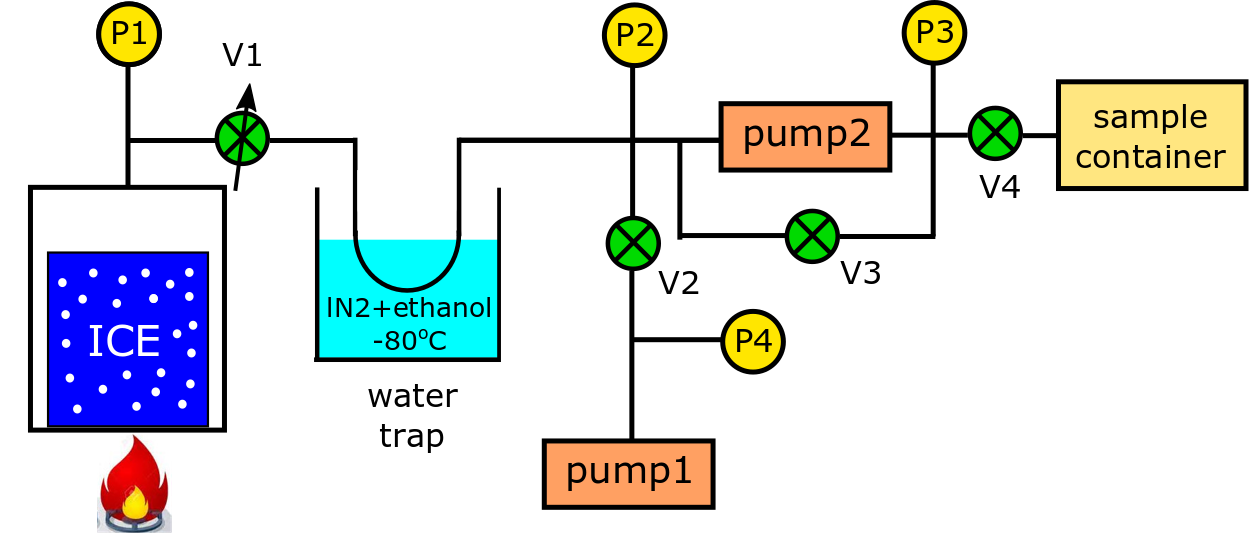}
\caption{Schematic of the setup for the extraction of air from the ice samples. P1-P4 are pressure gauges. V1-V4 denote valves in the pump line. Pump1 and pump2 are dry scroll vacum pumps for evacuating the system and compressing the released gas from the ice in the sample container (pump2). }
\label{fig:degassing_setup}
\end{figure}

A system for degassing the air from large (up to \SI{90}{kg}) glacier ice samples has been set up in the course of this study (Figure \ref{fig:degassing_setup}). The ice tank has a volume of \SI{140}{L} with \SI{50}{cm} inner diameter and \SI{60}{cm} in height. The lid is O-ring sealed and and has a window on top that allows visual monitoring of the state of the ice. Pump1 and pump2 are Edwards nXDS10i dry scroll vacuum pumps. They are both used to evacuate the system while pump2 is also used to compress the sample gas from the ice tank into the sample container.
 
V1 and V2 are O-ring sealed bellow valves. V3 and V4 are stainless steel welded bellow valves with 6mm tube press fittings. The opening range of V1 is used to control the flux of gas out of the tank. P1 and P3 are capacitance pressure gauges (\SI{100}{Pa}\hspace{0.05cm}–\hspace{0.05cm}\SI{200}{kPa} range). P1 is used to measure the pressure in the tank and P3 the pressure in the sample cylinder. P2 is a Baratron pressure gauge with an upper limit of \SI{1.1}{kPa}, which allows for leak testing and monitoring at low pressure conditions. P4 is a Pirani pressure gauge which is used to measure the pressure over a wide range from \SI{0.1}{Pa}\hspace{0.05cm}-\hspace{0.05cm}\SI{100}{kPa} when evacuating the system. The sample cylinders are made of stainless steel, have a volume of \SI{1.7}{L} and are sealed with V4. The water trap removes water vapor when evacuating the ice tank and compressing the air from the tank into the sample cylinder. The water trap is realised by a standard KF25 stainless steel bellow immersed into ethanol cooled by a commercial cooling device to \SI{-80}{\celsius}.
The air is extracted from the ice in three steps:

 (1) The ice is placed in the tank and the lid closed off. The residual atmospheric air is removed from the tank by pump1 and pump2 via the water trap which protects the pumps from the water vapor sublimating from the ice. After about 15 minutes a steady pressure of about \SI{100}{Pa} at P1 in the tank is established, which corresponds to the sublimation pressure of ice at around \SI{-20}{\celsius} (This is the storage temperature of the ice in the freezer). The tank is then pumped for another 15 minutes. Since during pumping the chamber is constantly being flushed by the water vapor from the sublimating ice, the remaining atmospheric gas in the container is decreased to a negligible level. 
 
(2) After removal of the atmospheric air, valve V1 is closed. The tank is heated with a gas stove until the ice is fully melted, a process that can be visually observed through a window in the lid of the tank. The melting process typically takes 60\hspace{0.05cm}-\hspace{0.05cm}\SI{90}{minutes} depending on the mass of the ice. While the ice is being melted, the sample cylinder and the rest of the system are evacuated with V2, V3 and V4 open.

(3) After the ice has fully melted the gas released from the ice is mostly in the headspace above the meltwater, since the volume of the tank is typically more than twice the volume of the water. Then, V2 and V3 are closed, V1 is opened (V4 remains open), and pump2 is used to compress the gas released from ice via the water trap to remove the water vapor. In order to keep the water vapor load low, dose valve V1 is regulated such that the pressure in the tank does not go below \SI{25}{mbar} which is close to the water vapor pressure at \SI{20}{\celsius}. After about 10 minutes of compression, the increase of the pressure in the sample cylinder (measured by P3) slows down. After a further 5 minutes, the sample cylinder is closed with valve V4. 
As the volume of the sample cylinder is known, the extracted amount of air can be obtained from the final pressure reading of P4 with an accuracy of about 5\%. The extraction efficiency of the degassing system has been measured by mixing degassed water with a known amount of air in the ice tank and compressing the air into the sample cylinder following Step (3). The measured extraction efficiency is $>$ 95\%.


\end{document}